\documentclass{appolb}
\usepackage{epsfig}
\usepackage{amssymb}
\newcommand{\Lepto}{\textsc{lepto}}
\newcommand{\Pythia}{\textsc{pythia}}

\begin{document}

\title{Diffractive Higgs production: \\
Soft Colour Interactions perspective
\thanks{Presented by N.\ T\^\i mneanu at the X International Workshop on 
Deep Inelastic Scattering (DIS2002), Cracow, 30 April - 4 May 2002}
}

\author{N.~T\^\i mneanu, R.~Enberg and G.~Ingelman 
\address{High Energy Physics, Uppsala University, 
Box 535, S-75121 Uppsala, Sweden}
}

\maketitle
\begin{abstract}

We briefly present the soft colour interaction models 
which are successful in reproducing a multitude of
diffractive hard scattering data,
and give predictions for diffractive Higgs production at
the Tevatron and LHC.
Only a few diffractive Higgs events may be produced at the Tevatron, but 
we predict a substantial rate at the LHC. 
\end{abstract}

Higgs production in diffractive hard scattering has been argued to be useful 
for Higgs discovery because of the lower hadronic background activity
in events with one or two rapdidity gaps and
leading protons. This especially holds for Higgs production in so-called 
double pomeron exchange (DPE) events, where the two beam protons survive the
collision, keeping a large fraction of the beam momentum, and where there is a
central system containing a Higgs. Another possibility is exclusive Higgs 
production, $p\bar p \to p\bar p H$, where the central system is just a Higgs
boson, and a missing mass method \cite{Albrow} can be applied.

Existing predictions \cite{diffr-Higgs-papers} of the cross sections for 
these processes vary by
several orders of magnitude, so the central question 
is whether the cross section is large enough.
In contrast to other models used for estimating the
diffractive Higgs cross section, our models have proven very successful in
reproducing experimental data on diffractive hard scattering processes both
from the $ep$ collider HERA and from $p\bar{p}$ collisions at the
Tevatron \cite{SCI-TEV}. This puts us in a better
position to answer the question whether the diffractive Higgs channel is a
feasible one at the Tevatron and at LHC \cite{SCI-higgs}.

%---------------------
\begin{figure}[t]
\begin{center}
\epsfig{width= 0.6\columnwidth,file=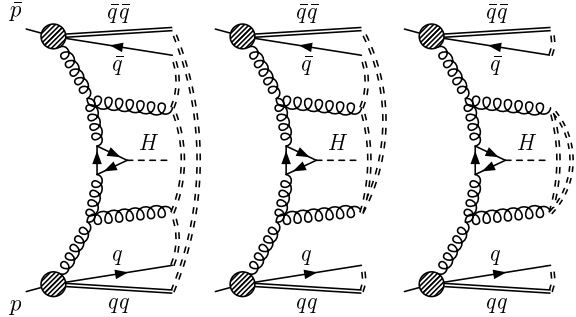,clip=}
\caption{\small\it Higgs production in $p\bar{p}$ collisions with string topologies
(double-dashed lines) before and after soft colour interactions in the SCI or
GAL model, resulting in events with one or two rapidity gaps 
(leading particles).}\vspace*{-8mm}
\label{pp-higgs}
\end{center}
\end{figure}
%---------------------

The soft colour interaction (SCI) model \cite{SCI} and the generalized area law
(GAL) model \cite{GAL} were developed under the assumption that soft colour exchanges
give variations in the topology of the confining colour string-fields such that different final
states could emerge after hadronization, \eg \ with and without rapidity gaps or
leading protons.

Both models are implemented in the Lund Monte Carlo programs \Lepto
\ \cite{Lepto} for deep inelastic scattering and \Pythia \ \cite{Pythia} for
hadron-hadron collisions. The hard parton level interactions are given by
standard perturbative matrix elements and parton showers, which are not altered
by the softer non-perturbative effects. The SCI model then applies an explicit
mechanism where colour-anticolour (corresponding to non-perturbative gluons) can
be exchanged between the emerging partons and hadron remnants. 
The GAL model, similar in spirit, is formulated in terms of
interactions between the strings. The soft colour exchanges
between partons or strings change the colour topology resulting in another string
configuration (Fig.~\ref{pp-higgs}). 
The probability for such an exchange is taken to be a constant
phenomenological parameter in the SCI case, while for GAL  the probability 
for two strings to interact is dynamically varying, favoring ``shorter'' strings
and suppressing `longer' ones. The only parameter entering the models
has its value determined from the HERA rapidity gap data and then is kept fixed. 

\begin{figure}[t]
\begin{center}
\epsfig{width= 0.55\columnwidth,file=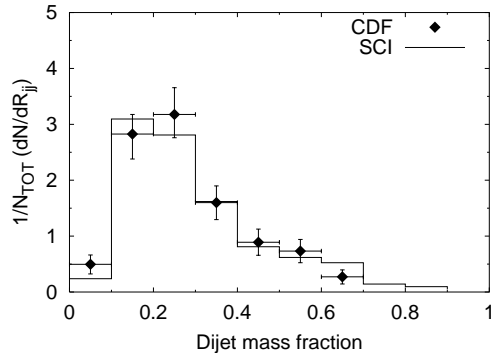,clip=}
\caption{\small\it Distribution of the 
dijet mass fraction (ratio of
dijet mass to total central system mass)
in DPE events at the Tevatron, CDF data \cite{CDF-DPE} compared to 
SCI model.}\vspace*{-8mm}
\label{dijet-mass}
\end{center}
\end{figure}

The SCI and GAL models give different diffractive hard scattering processes by
simply choosing different hard scattering subprocesses in \Pythia. Rapidity gap
events containing a $W$, a dijet system or bottom quarks are found to be in
agreement with Tevatron data \cite{SCI-TEV}. Diffractive events with a leading proton,
or two leading protons \cite{CDF-DPE}, are also well described \cite{SCI-TEV}. In particular, 
the cross sections for dijets in DPE events obtained from the models
agree with the CDF data \cite{CDF-DPE}, as do more exclusive quantities, such as
the dijet mass fraction (see Fig.~\ref{dijet-mass}). Let us emphasize that the
dynamics of this process is similar to the DPE Higgs process, and this has been
advocated as a testing ground for different models aiming at describing
diffractive Higgs.

%---------------------
%---------------------

The predictive power of the models has also been tested, since we were able to predict
correct ratios for production of $J/\psi$ associated with gaps at the Tevatron \cite{SCI-TEV}.
It is remarkable how through the same soft colour mechanism,
two different soft phenomena arise in the same event, namely a rapidity gap and
turning a colour octet $c\bar{c}$ pair into a colour singlet producing $J/\psi$.
Furthermore, our predicted ratios for diffractive $Z$ production \cite{SCI-TEV} seem to
be in very good agreement with those recently found by D\O~\cite{Z}. 

The properties of the Higgs boson in the Standard Model are fixed, except for
its mass. The present lower limit is 114.1 GeV and $\chi^2$ fits to high
precision electroweak data favors $m_H < 212$ GeV \cite{Higgslimit}. The 
latest LEP data give an indication ($\sim 2.1\, \sigma$) of a Higgs with a mass
of 115.6 GeV \cite{Higgsevidence}.

Higgs production at
the Tevatron and the LHC can proceed through many subprocesses,
which are included in \Pythia \ version 6 \cite{Pythia}. The dominant one is
$gg\to H$, which accounts for 50\% and 70\% of the cross section (for
$115<m_H<200$ GeV) at the Tevatron and LHC, respectively. In this process, see
Fig.~\ref{pp-higgs}, the gluons couple to a quark loop with dominant
contribution from top due to its large coupling to the Higgs. Other production
channels are also considered. The overall cross sections are obtained by folding the
subprocess cross sections with the parton density distributions. 
%

%---------------------
\begin{table}[t]
\caption{\small\it Cross sections at the Tevatron and LHC for Higgs in single diffractive
(SD) and DPE events, using leading proton or rapidity gap definitions, as well
as relative rates (SD/all and DPE/SD) and number of events, obtained from the
soft colour exchange models SCI and GAL.\label{tab-higgs}}
\small{
\begin{center}
\begin{tabular}{lcccc}
\hline
\hline
               &  \multicolumn{2}{c}{Tevatron}                       &  \multicolumn{2}{c}{LHC} \\
$m_H=115$~GeV  &  \multicolumn{2}{c}{$\sqrt{s}=1.96$~TeV}  
               &  \multicolumn{2}{c}{$\sqrt{s}=14$~TeV} \\
               &  \multicolumn{2}{c}{${\cal L}=20~\mbox{fb}^{-1}$}   
               &  \multicolumn{2}{c}{${\cal L}=30~\mbox{fb}^{-1}$} \\
\hline
$\sigma [\mbox{fb}]$ Higgs-total    & \multicolumn{2}{c}{600}        & \multicolumn{2}{c}{27000} \\
\hline 
\hline
               & SCI & GAL & SCI & GAL \\
\hline 
\multicolumn{5}{l}{Higgs in single diffraction:}\\
$\sigma \; [\mbox{fb}]$ leading-p   & 1.2 & 1.2 & 190 & 160 \\
$\sigma \; [\mbox{fb}]$ gap         & 2.4 & 3.6 & 27 & 27   \\
$R \; [\%]$ leading-p               & 0.2 & 0.2 & 0.7 & 0.6 \\
$R \; [\%]$ gap                     & 0.4 & 0.6 & 0.1 & 0.1 \\
\# H + leading-p                    & 24 & 24 & 5700 & 4800 \\
$\hookrightarrow$ \# H $\rightarrow \gamma\gamma$  & 0.024 & 0.024 & 6 & 5 \\
\hline 
\multicolumn{5}{l}{Higgs in DPE:}\\
$\sigma \; [\mbox{fb}]$ leading-p's & $1.2 \cdot 10^{-4}$   & $2.4 \cdot 10^{-4}$   & 0.19 & 0.16 \\
$\sigma \; [\mbox{fb}]$ gaps        & $2.4 \cdot 10^{-3}$   & $7.2 \cdot 10^{-3}$   & $2.7 \cdot 10^{-4}$   & $5.4 \cdot 10^{-3}$ \\
$R \; [\%]$ leading-p's             & 0.01                  & 0.02                  & 0.1                   & 0.1 \\
$R \; [\%]$ gaps                    & 0.1                   & 0.2                   & 0.001                 & 0.02 \\
\# H + leading-p's                  & 0.0024                & 0.0048                & 6                     & 5 \\
\hline
\hline
\end{tabular}
\end{center}}\vspace*{-8mm}
\end{table}
%---------------------

After the standard parton showers in \Pythia, SCI or GAL is applied, giving a total sample
of Higgs events, with varying hadronic final states. Single
diffractive (SD) Higgs events are selected using one of two criteria: (1) a
leading (anti)proton with $x_F>0.9$ or (2) a rapidity gap in $2.4<|\eta|<5.9$
as used by the CDF collaboration. Applying the conditions in both hemispheres
results in a sample of DPE Higgs events. The resulting cross sections and
relative rates are shown in Table \ref{tab-higgs}. The results have an
uncertainty of about a factor two related to details of the hadron remnant
treatment and choice of parton density parameterization.

The cross sections at the Tevatron are quite low in view of the luminosity to
be achieved in Run~II. Higgs in DPE events are far below an observable rate. For
$m_H=115$ GeV, only tens of single diffractive Higgs events are predicted. Only
the most abundant decay channel, $H\to b\bar{b}$, can then be of use and a very
efficient $b$-quark tagging and Higgs reconstruction is required. The
conclusion for the Tevatron is thus that the advantage of a simplified
reconstruction of the Higgs in the cleaner diffractive events is not really
usable in practice due to a too small number of diffractive Higgs events being
produced. 

\begin{figure}
\begin{center}
\epsfig{width= 0.55\columnwidth,file=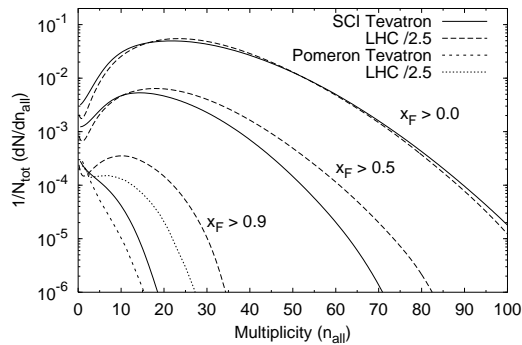}
\caption{\small\it Multiplicity (for LHC divided by 2.5) in the region $2.4<|\eta|<5.9$
in the hemisphere of a leading
proton with the indicated minimum $x_F$, for Higgs events from the SCI and the
pomeron models.}\vspace*{-8mm}
\label{fig-multiplicity}
\end{center}
\end{figure}

In contrast, the high energy and luminosity available at the LHC facilitate a
study of single diffractive Higgs
production, where also the striking $H\to \gamma \gamma$ decay should be
observed. Also a few DPE Higgs events may be observed. The quality of a
diffractive event changes, however, at LHC energies. Besides the
production of a hard subsystem and one or two leading protons, the energy is
still enough for populating forward detector rapidity regions with particles.
As seen in Fig. \ref{fig-multiplicity}, the multiplicity of
particles is considerably
higher at the LHC, compared to the Tevatron. The requirement of
a ``clean'' diffractive Higgs event with a large rapidity gap in an observable
region cannot be achieved without paying the price of a lower cross section.
Requiring gaps instead of leading
protons gives a substantial reduction in the cross section, as seen in 
Table~\ref{tab-higgs}. Note that the high luminosity mode of LHC cannot be used, 
since the resulting pile-up of events would destroy the rapidity gaps.

%---------------------
%---------------------

In conclusion, the soft colour interactions models predict a rate of diffractive 
Higgs events at the Tevatron  which is too low to be useful. However, LHC should facilitate 
studies of Higgs in single diffraction and the observation of some DPE events with a Higgs boson.

%---------------------------------------------------------------

\end{document}